\documentclass[pdflatex,sn-aps]{sn-jnl}

\usepackage{graphicx}%
\usepackage{multirow}%
\usepackage{amsmath,amssymb,amsfonts}%
\usepackage{amsthm}%
\usepackage{mathrsfs}%
\usepackage[title]{appendix}%
\usepackage{xcolor}%
\usepackage{textcomp}%
\usepackage{manyfoot}%
\usepackage{booktabs}%
\usepackage{algorithm}%
\usepackage{algorithmicx}%
\usepackage{algpseudocode}%
\usepackage{listings}%

\begin{document}

\title[Article Title]{Statistical methods for resolving poor uncertainty quantification in machine learning interatomic potentials}


\author*[1]{\fnm{Emil} \sur{Annevelink}}\email{eannevel@andrew.cmu.edu}

\author*[1]{\fnm{Venkatasubramanian} \sur{Viswanathan}}\email{venkvis@cmu.edu}

\affil[1]{\orgdiv{Mechanical Engineering}, \orgname{Carnegie Mellon University}}


\abstract{
Machine learning interatomic potentials (MLIPs) are great candidates to surrogate quantum mechanics evaluations for ab-initio molecular dynamics simulations due to their ability to reproduce the energy and force landscape within chemical accuracy at four orders of magnitude less cost.
However, producing robust MLIPs for a target material system remains a challenge due to the need for sampling rare events and accurately labeling configurations that are distinct from the training dataset.
While developing uncertainty quantification tools for MLIPs is critical to build production MLIP datasets using active learning, only limited progress has been made and the most robust method, ensembling, still shows low correlation between high error and high uncertainty predictions.
Furthermore, in active learning workflows a cutoff is needed to determine what corresponds as high error and high uncertainty, with the definition of high uncertainty being different for each uncertainty model.
Here we develop a rigorous method rooted in statistics for determining the cutoff between high and low error.
The statistical cutoff illuminates that a main cause of the poor uncertainty quantification performance is due to the machine learning model already describing the entire dataset and not having any datapoints with error greater than the statistical error distribution.
Second, we extend the statistical analysis to create an interpretable connection between the error and uncertainty distributions to predict an uncertainty cutoff separating high and low errors.
We showcase the statistical cutoff in active learning benchmarks on two datasets of varying chemical complexity for three common uncertainty quantification methods: ensembling, sparse Gaussian processes, and latent distance metrics and compare them to the true error and random sampling.
We show that the statistical cutoff is generalizable to a variety of different uncertainty quantification methods and protocols and performs similarly to using the true error.
We conclude by discussing how the statistical cutoff enables accurate classification of high error datapoints even for poorly calibrated models.
This enables using significantly lower cost uncertainty quantification tools such as sparse gaussian processes and latent distances compared to ensembling approaches for generating MLIP datasets at a fraction of the cost.
}




\maketitle

\section{Introduction}

Machine Learning Interatomic Potentials (MLIPs) have shown tremendous potential to surrogate energy and force calculations from quantum mechanical simulations to provide a significant improvement to the accuracy cost trade-off between density functional theory (DFT) and classical potentials \cite{BehlerGeneralized2007, wenHybrid2019, musaelianScalingLeadingAccuracy2023}.
Different machine learning architectures have emerged including descriptor based \cite{BehlerGeneralized2007}, Hamiltonian \cite{greydanus2019hamiltonian}, continuous filter \cite{schutt2018schnet} and equivariant \cite{batznerEquivariantGraphNeural2022} descriptions.
Each architecture has a particular accuracy-cost trade-off in addition to its data efficiency based on its representation power of the dataset it is trained on.
Benchmark datasets such as MD17 \cite{chmielaMachineLearningAccurate2017} have been critical to architecture development, but as the architectures improved, the need for being able to develop production datasets became increasingly clear.

Ab-initio molecular dynamics (AIMD) simulations can be utilized to generate an initial dataset of samples.
When trained on AIMD datasets, MLIPs are able to sample larger supercells for longer simulations to generate better statistics to reveal emergent physics \cite{winterSimulationsMachineLearning2022}.
However, AIMD does not always sample the entire phase space, so when MLIPs sample a region of unknown phase space, they become unstable \cite{schwalbe-kodaDifferentiableSamplingMolecular2021,tanSinglemodelUncertaintyQuantification2023}.
Importantly, AIMD cannot scale to build datasets that sample rare events.

Active learning workflows, where MLIPs self-identify gaps in their training data distribution using an uncertainty estimator have been utilized to sample rare-events and produce high-accuracy MLIPs.
Active learning workflows have been developed for numerous uncertainty models including ensembling \cite{wang2018deepmd,zhuFastUncertaintyEstimates2022}, dropout Monte Carlo \cite{wenUncertaintyQuantificationMolecular2020a}, and sparse Gaussian Processes \cite{vandermauseOntheflyActiveLearning2020,vandermauseActiveLearningReactive2022a,zhuFastUncertaintyEstimates2022}.
A key feature in the active learning workflows are two cutoffs.
A minimum cutoff determines the uncertainty threshold for when you are accurately surrogating DFT, and a maximum cutoff determines a threshold for datapoints too far outside of the training dataset.
Determining the cutoffs is challenging due to different definitions for the uncertainty between each uncertainty method.
Recently, work has been done to address the performance of uncertainty quantification in machine learning models to improve the identification of relabeling points \cite{tanSinglemodelUncertaintyQuantification2023,rensmeyerHighAccuracyUncertaintyAware2023}.
Many of these works conclude by acknowledging that there is not a clear understanding for how to evaluate the performance of an uncertainty model.
So, while they can show that active learning improves the stability of a MLIP, there is no clarity on the correct evaluation metrics and therefore little indication on how to improve the performance of uncertainty quantification.

Here we develop a robust method for linking a desired error cutoff to uncertainty cutoffs for a generic uncertainty model to automatically determine interpretable minimum and maximum uncertainty cutoffs.
We start by building a theoretical understanding of MLIP errors from a statistical viewpoint.
We use this understanding to build a statistical cutoff for the error to determine if it is within the training/validation dataset or not.
We then apply this same theoretical basis to the uncertainty quantification to establish an uncertainty cutoff.
These cutoffs create a theoretical underpinning for the minimum uncertainty cutoff for any UQ method and show that only in the highly sparse data regime are there any points outside of the training distribution.
We further integrate the statistical cutoff into an active learning workflow to compare the performance of difference UQ strategies for MLIPs.
We find that using the statistical cutoffs works makes all the tested UQ strategies robust even with poor calibration curves.
We posit that most efforts to benchmark the uncertainty predictions fail due to their being few, if any, uncertain datapoints in the benchmark.
This is especially true for data-efficient equivariant models which we utilize throughout this paper\cite{batznerEquivariantGraphNeural2022} built on top of the E3NN pytorch package \cite{e3nn,e3nn_paper}, but believe that the statistical methods apply to descriptor based MLIPs as well.

\section{Errors in Machine Learning Interatomic Potentials}

\begin{figure*}[t]
    \centering
    \includegraphics[width=0.9\textwidth]{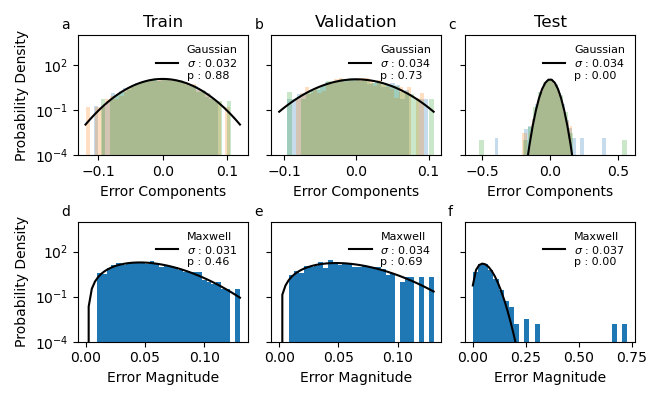}
    \caption{Distribution of x, y, z force error components in blue, orange, green respectively (a-c) and magnitudes (d-f) for the training (a/d), validation (b/e), and testing (c/f) splits of a lithium benchmark dataset sampled at 600K in an NVT ensemble. Further dataset details are found in the SI. The error components and magnitudes are fit to Gaussian and Maxwell distributions respectively, with their standard deviations and p values from KS-tests reported on the figures.}
    \label{fig:error_definition}
\end{figure*}

The theoretical distribution for trained deep neural network predictions is a Gaussian \cite{roberts_yaida_hanin_2022}.
The statistical thermodynamics analysis utilizes perturbation theory to show that for highly over-parameterized regimes a deep neural network model will train well and match the distribution mean to the data labels.
Since the model predictions are distributed according to a Gaussian, the expected error distribution of these models is Gaussian.

The label most often used to train MLIPs is force matching, where the three--x,y,z--force components are fit using a loss function.
Force matching is used due to its per-atom definition enabling the use of spatial locality to form the underlying graph / neighborlist for the interatomic potentials.
Since the force prediction is a vector that is equivariant under symmetry operations of the atomic configuration \cite{fuchsSETransformers3D2020}, the force errors are also equivariant due to the true and predicted forces transforming identically with symmetry operations.
However, the per-atom error should be an invariant quantity that does not transform according to symmetry operations.
This is overcome by using the force error magnitude, an invariant descriptor for per-atom errors, with both the MSE or RMSE being common vector-to-scalar transformations.

The force errors for benchmark dataset 1 for which further details can be found in the SI are plot in Figure \ref{fig:error_definition}(a-c).
The x, y, and z components are plot seperately as blue, orange, and green histograms respectively.
For all the components, the component force error distribution is approximated with a Gaussian distribution for the train, validation, and testing dataset.
The Gaussian center is fixed at error$=0$, and the standard deviation is fit.
Using the standard deviation of the Gaussian, a Kolmogorov-Smirnov (KS) test is used to check the validity of the fit.
The null hypothesis is that the data has a Gaussian distribution, which for both the training and validation set is rejected with p-values of 0.88 and 0.73 respectively.
However, for the test set, the null hypothesis is not rejected due to the presence of large errors far away from the central distribution.

The expected distribution for the sum of squared random Gaussian variables is either a $\chi^2$ distribution, which becomes a $\chi$ distribution if the square root is taken.
In three dimensions, the $\chi$ distribution is the well known Maxwell distribution with a single degree of freedom, the standard deviation.
In Figure \ref{fig:error_definition}(d-f), the magnitude of the errors corresponding to the component errors is plot.
A KS-test is similarly done to evaluate the goodness of fit for Maxwell distributions to the error magnitudes.
Again, for the train and validation sets, the null hypothesis is rejected and Maxwell distributions fit well, but for the test set, the null hypothesis cannot be rejected and similar to the Gaussian distribution, large error magnitudes exist, where the Maxwell distribution predicts very low probability density.

With the insight that the training and validation distributions are Maxwell distributions, we developed a confidence interval to create a theoretical cutoff for samples that lay outside of the Maxwell distribution for a given dataset size.
The confidence interval was constructed such that the cumulative distribution function would be equal to $1-1/N$, where N is the size of a sampling dataset.
The error magnitude distribution and cutoff are shown in Figure \ref{fig:uncertainty_definition}(a) for the validation dataset and in Figure \ref{fig:uncertainty_definition}(d) for the sampling dataset.
As can be seen, the cutoff distinguishes the transition from being described by the Maxwell distribution to not being described by the Maxwell distribution as defined by the similarity of the probability distributions.

\section{Uncertainty Prediction in Machine Learning Interatomic Potentials}

\begin{figure*}[t]
    \centering
    \includegraphics[width=0.9\textwidth]{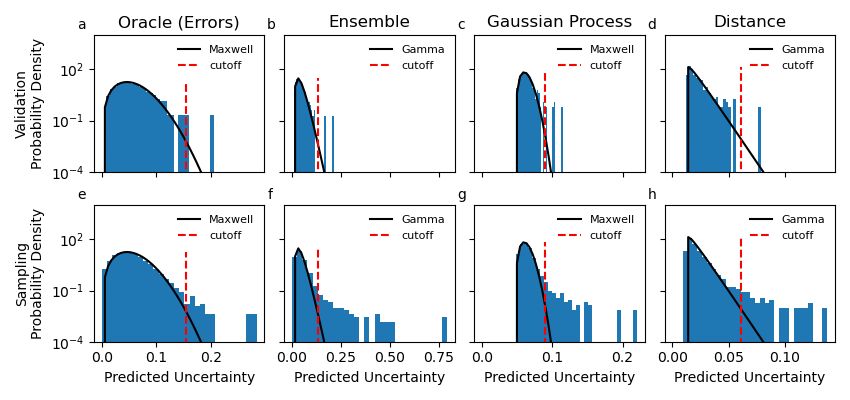}
    \caption{The uncertainty predictions for the validation dataset (a-c) and sampling dataset (d-f) of dataset 1 predicted using an oracle (the errors), an ensemble of models, sparse Gaussian process, and a latent distance model. The model PDF and statistical cutoff of the highest p-value according to a KS-test are also plotted.}
    \label{fig:uncertainty_definition}
\end{figure*}

In active learning, the error is not known for the sampling distribution a-priori.
The error needs to be predicted using an uncertainty classifier.
The most common uncertainty method in MLIPs is to create an ensemble of models initialized with different random weights \cite{zhangDPGENConcurrentLearning2020,wenUncertaintyQuantificationMolecular2020a}.
The second most common method is to use sparse Gaussian Processes to reduce the cost of training up to ten MLIPs in an ensemble\cite{zhuFastUncertaintyEstimates2022,vandermauseOntheflyActiveLearning2020}.
Here we also introduce the use of a distance based classifier to predict the standard deviation of a Maxwell distribution\cite{janetQuantitativeUncertaintyMetric2019}.
These methods have been described extensively elsewhere, and are reviewed in the SI.
Here we discuss how to develop statistical cutoffs for each method to be used in an active learning workflow.

The uncertainty cutoffs are determined by fitting a distribution to the predicted uncertainties of each of the methods. 
In addition to the theoretical Maxwell distribution, we also test Gamma and generalized Gamma distributions.
A Gamma distribution becomes a Maxwell distribution when using $3/2$ degrees of freedom and as the name suggests, a generalized Gamma distribution has one more shape parameter than a Gamma distribution.
Although for each of these uncertainty methods, the theoretical distribution could be worked out, here we only use the insight that the uncertainty distribution should follow a statistical distribution and utilize a set of sufficiently descriptive distributions to find one that matches the predicted uncertainties above a 95\% confidence level as given by a KS-test.

The uncertainty cutoff is determined using the same approach as the error cutoff, where the confidence interval is defined by when the cumulative distribution function is equal to $1-1/N$, where N is the sampling dataset size.
The uncertainty distributions predicted by ensembling, sparse Gaussian process, and latent distance are shown for the validation and sampling datasets in Figure \ref{fig:uncertainty_definition}.
For all the uncertainty methods, a number of points are above the cutoff for the sampling dataset, including the true errors.

The ability for the uncertainty cutoffs to accurately identify high error points can be determined using classification metrics such as the precision, recall, and F1 score.
The precision is defined as
\begin{equation}
    Precision = \frac{True Positive}{True Positive + False Positive}
\end{equation}
and determines the percentage of points that are labeled uncertain that actually have high error.
Similarly, the recall is defined as
\begin{equation}
    Recall = \frac{True Positive}{True Positive + False Negative}
\end{equation}
and defines the percentage of points that have high error that are labelled as uncertain.
The ideal classifier would have unity precision and recall, in practice, however, the precision and recall are in competition and need to be balanced.
The F1 score is a metric that reports a specific balance of these two metrics and is defined as
\begin{equation}
    F1 = \frac{2*Precision*Recall}{Precision+Recall}.
\end{equation}

\begin{figure*}[t]
    \centering
    \includegraphics[width=0.8\textwidth]{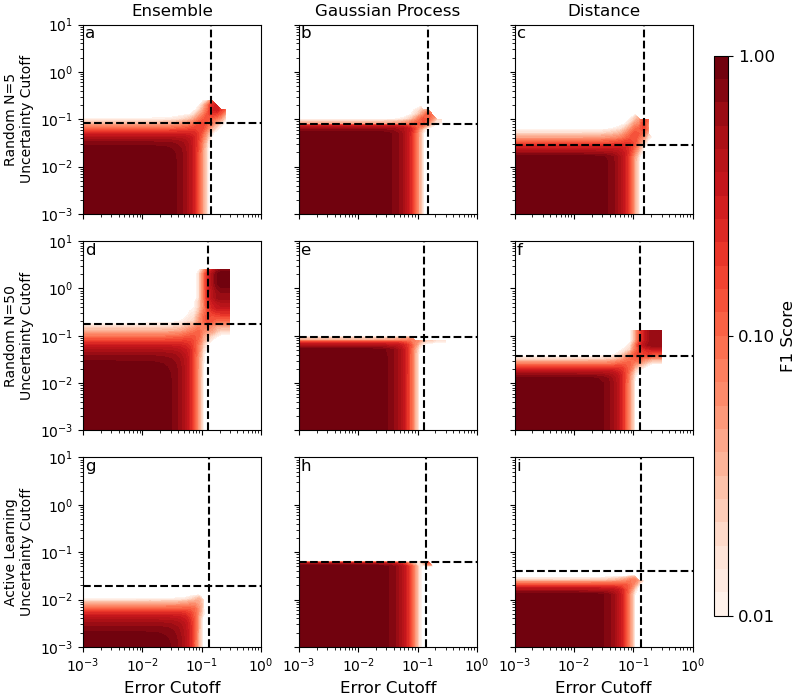}
    \caption{The precision, recall, and F1 scores for a sweep of error and uncertainty cutoffs using the uncertainty predicted from the standard deviation of an ensemble of models. (a-c) are for the first active learning index with a total dataset size of XX and (d-f) are for the last active learning index with a final dataset size of YY.}
    \label{fig:prec_rec_f1}
\end{figure*}

The F1 scores for the three uncertainty methods on Dataset 1 for different error and uncertainty cutoffs are shown in Figure \ref{fig:prec_rec_f1} for 5 and 50 randomly sampled points and samples selected using active learning with dataset sizes of 49, 33, and 27 respectively for ensembling, Gaussian Process, and distance.
For the randomly sampled data, the MLIP splits are the same for each of the methods, and the underlying MLIP for the Gaussian Process and Distance uncertainty method is identical.
The colormap reports the F1 score for a complete sweep of error and uncertainty cutoffs, and the statistical cutoffs for both the error and uncertainty are identified as dotted black lines.
The sweeps of the precision and recall are provided in the SI as reference. 
As is expected, for low error and uncertainty cutoffs, the F1 score is near unity as the majority of datapoints are labelled as uncertain.
Furthermore, along each axis, the F1 score rapidly decreases to zero but for high error and uncertainty cutoffs there is a bridge that connects the interpolation region in the bottom left and the extrapolation region in upper right.


For the ensembles trained on randomly sampled data in Figure \ref{fig:prec_rec_f1}(a,d), the statistical cutoffs are located at this bridge that extends into the high error and uncertainty region.
In combination with the parity and calibration plots in the SI, we can confirm that the ensembling method outperforms the Gaussian Process and Distance methods at identifying uncertain datapoints for this dataset.
When examining the regions separated by the bridges, namely Figure \ref{fig:prec_rec_f1}(a,c,d,f) a higher F1 score is achieved past the cutoff regions showing that a higher classification accuracy cutoff exists than this statistical cutoff. 
However, that the statstical cutoffs are always located at the bridge to this higher classification accuracy region signifies that the statistical cutoffs are a conservative estimation of the ideal cutoff and represent a boundary between two regions of decreasing and increasing F1 scores.
Furthermore, we can compare the performance of the Gaussian Process and Distance methods by comparing Figure \ref{fig:prec_rec_f1}(e,f), which use the same MLIP but diferent uncertainty classifiers.
The poor calibration of the Gaussian Process model appears as the absence of a bridge in Figure \ref{fig:prec_rec_f1}(e), which is present in Figure \ref{fig:prec_rec_f1}(f)  as the Gaussian Process is not labelling the high error points that the Distance method is labelling as uncertain.

Next, we compare the F1 score distributions of the randomly sampled datasets to the active learning dataset.
The active learning dataset was seeded with five randomly sampled points and the dataset was increased iteratively with the most uncertain points until no uncertain points remained.
A key differentiator between the active learning and randomly sampled datasets is that the F1 score monotonically decreases away from the origin.
After the active learning iterations, there are no longer any points outside of the statistical cutoffs and all datapoints are in the interpolation regime of the dataset.
For the ensemble and distance methods in Figure \ref{fig:prec_rec_f1}(g,i), the F1 score decreases over a larger range of cutoffs as compared to the Gaussian Process model in Figure \ref{fig:prec_rec_f1}(h).
This is due to the narrower range of the Gaussian Process predicted uncertainties, which is also observed in the parity plots in the SI.
We note that although the error cutoffs are nearly uniform across all the tests, the predicted uncertainty cutoffs vary between models but also with different dataset sizes.
This shows the flexibility of using distributions to correspond between error and uncertainty.

By comparing the F1 scores across cutoffs, we can also put the results of Zhu et al. \cite{zhuFastUncertaintyEstimates2022} into perspective, where they identified a monotonically decreasing precision*recall.
This means that their MLIPs were already describing their entire sampling dataset and therefore could not observe the bridge to increasing F1 scores.
Therefore, benchmarking uncertainty models for MLIPs might benefit from utilizing data-starved models that cannot describe the entire dataset, which is especially true for data-efficient equivariant models that Zhu et al. were utilizing. 
For further understanding of the relationship between the errors and uncertainties, parity plots and calibration curves for each uncertainty method as well as the precision and recall curves are provided in the SI.

\section{Active Learning Tests}

\begin{figure*}
    \centering
    \includegraphics[width=0.9\textwidth]{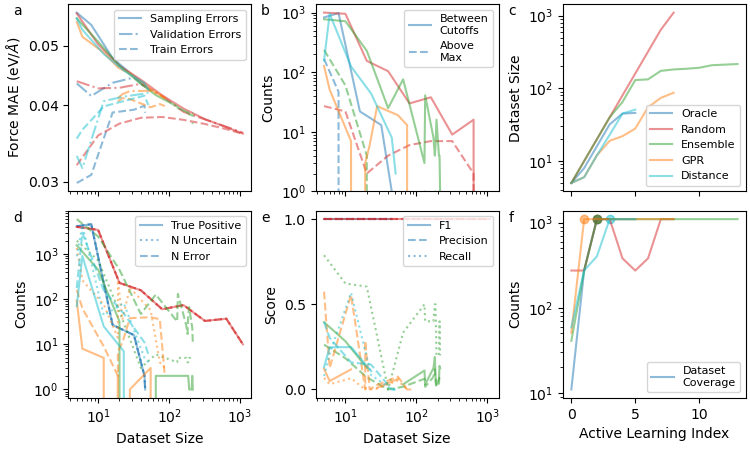}
    \caption{Performance of ensembling, Gaussian Process, and Distance uncertainty models compared to an oracle and random sampling on a single component benchmark dataset of melted lithium. Concatenation of Datasets 2 and 3 in the SI. The metrics are presented versus dataset size (a,b,d,e) and active learning index (c,f).}
    \label{fig:Li_sequential_combined}
\end{figure*}

We test the ability of different uncertainty quantification methods to successfully build MLIP datasets using the statistical cutoffs.
We aim to reproduce a realistic active learning workflow that builds a dataset from sampling phase space using molecular dynamics.
The first 100 samples are uniformly sampled from a 1ps simulation, and the next 1000 samples are uniformly sampled from a 10ns simulation for a total dataset size of 1100.
Two different benchmarks are utilized.
The first focuses on a liquid single chemical species (lithium) dataset and the second on a multi-component (Li-Mg), multi-phase (solid-liquid), and multi-lattice (BCC-HCP) dataset to increase the data complexity.
The details of these datasets are found in the SI.
The first corresponds to concatenating datasets 2 \& 3 and the second uses datasets 4-13.
Since the entire dataset is already calculated, to reproduce the instability of using MLIPs to sample phase space, a force error cutoff is established that corresponds to a change of velocity that would move an atom more than 10\% of the nearest neighbor distance or $\approx0.3\AA$ or roughly 1eV/$\AA$.
No datapoint can be added to the MLIP dataset that is greater than the first point that has an uncertainty above this upper cutoff such that the MLIP dataset for each method is created by sampling the entire dataset sequentially using active learning.

In addition to the three uncertainty methods, we include an upper and lower performance bound.
The best performance for active learning is found by defining the predicted uncertainty as the true error.
The worst performance active learning is not using any uncertainty information and randomly sampling from the sampled dataset.
Expectedly, the random sampling learns very slowly and is therefor stopped early.

The results of the first active learning test is shown in Figure \ref{fig:Li_sequential_combined}.
The active learning metrics are the mean of the error magnitudes, the number of points in the sampling dataset between the cutoffs and above the maximum cutoff, the fraction of samples identified as true positives, high uncertainty, and high error, the precision, recall, and F1 scores, and the dataset size and coverage.

The means of the error magnitudes for the training, validation, and sampling distributions are shown in Figure \ref{fig:Li_sequential_combined}(a) against the dataset size.
At first glance, each of the methods perform identically when looking at the ``Sampling Errors''.
However, we can see a significant difference when comparing the validation and training errors between each method.
While the validation error decreases for the random sampling, it increases for the oracle until it intersects with the ``Sampling Errors'' curve.
The Gaussian Process, distance, and ensemble validation means increase.
This is an indication that samples with high error are being included in the validation dataset and the uncertainty approaches are capturing the high error data points.
Furthermore, the error magnitudes of the training dataset increase as the amount of data increases.
This is due to increased dataset complexity as more diverse data is added to the training distribution.

The active learning performance is further studied by visualizing the number of datapoints between the minimum and maximum cutoffs and those greater than the maximum cutoff against dataset size in Figure \ref{fig:Li_sequential_combined}(b).
The oracle and random sampling provide an upper and lower bound to the uncertainty methods.
While the oracle, random sampling, and distance model decrease fairly consistently, the Gaussian process model and ensemble models are more erratic, where the number of uncertain points does not decrease monotonically.
The erratic nature will be discussed further in conjuction with Figure \ref{fig:Li_sequential_combined}(d,e).
Still, the uncertainty models all perform as expected between the best and worst limit.

Next, the classification statistics covered by the true positive, number of uncertain and high error points, and Recall, Precision, and F1 scores against dataset size are in Figure \ref{fig:Li_sequential_combined}(d,e).
Immediately, the stochastic nature of training MLIPs appears, where the number of high error points is dramatically different for the initial smallest dataset size.
Therefore, while the number of uncertain and high error data points for the oracle and random sampling decrease monotonically \ref{fig:Li_sequential_combined}(d), they do not bound the number of error points due to significantly different initial counts of datapoints with high error.
The erratic nature of the training makes the plots challenging to interpret.
Still, we can see that the overlap between the number of uncertain points and error points is perfect by definition for the oracle and random sampling, but the overlap for the approximate uncertainty models is far from ideal as the number of error points and number of uncertain points diverge strongly.
Interestingly, while the number of error points is very low initially for the Gaussian process, it increases as the dataset complexity increases.
This is in stark contrast to the ensemble model, where the number of error points mirrors the randomly sampled model up until 100 datapoints before it decreases.
The mismatch between the error and uncertainty distributions can also be seen in the precision, recall, and F1 scores.
The precision, recall, and F1 scores are nearly identical for the distance metric, but the recall and precision for ensembling strongly oscillate with dataset size, while the Gaussian Process losses all True Positive datapoints early preventing any analysis there.
The oracle uncertainty method (also used for Random sampling) gives an expected Precision, Recall, and F1 score of unity in \ref{fig:Li_sequential_combined}(e).
As a whole the F1 decreases with increasing dataset size, which is expected since the task is becoming more challenging with fewer and fewer points with high error.

The data efficiency and sampling stability is reported in Figure \ref{fig:Li_sequential_combined}(c,f).
The linear slope in Figure \ref{fig:Li_sequential_combined}(c) comes from limiting the dataset size to no more than double.
Any fall-off from doubling indicates the lack of uncertain points for relabelling.
As expected, the oracle is the most data efficient method, where it outperforms the approximate methods in correctly identifying the last remaining high error points.
The path for achieving a final dataset size differs for all the methods.
The oracle, ensembling, and distance model all double their dataset sizes until a final plateau, whereas the Gaussian process model increases its dataset size in two phases mirroring the lack of high error points in the initial active learning iterations.
The final dataset size for the distance model is nearly identical to the oracle dataset size.
The ensembling model on the other hand requires more than double the datapoints and many more active learning iterations.
Considering it also requires training multiple MLIPs each iteration, the higher accuracy of the uncertainty predictions comes at a higher cost.

\begin{figure*}
    \centering
    \includegraphics[width=0.9\textwidth]{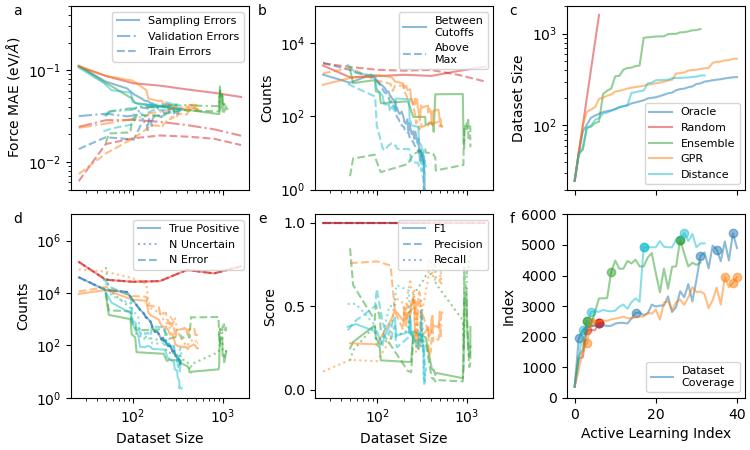}
    \caption{Performance of ensembling, Gaussian Process, and Distance uncertainty models compared to an oracle and random sampling on a multi-phase, multi-component benchmark dataset of solid and melted lithium and magnesium mixtures. Five separate compositions are trained on simultaneously representing Datasets 4-13 in the SI. The metrics are presented versus dataset size (a,b,d,e) and active learning index (c,f). In (f), the scatter points represent the indices when individual datasets reach their maximum value.}
    \label{fig:LiMg_sequential_combined}
\end{figure*}

The second active learning test assesses the performance of the statistical cutoff and different active learning methods on a chemically complex dataset. 
The second active learning test spans datasets 4-13 as defined in the SI and covers varying compositions of lithium and magnesium, solid and liquid phases, and BCC and HCP lattices.
The results are plotted in Figure \ref{fig:LiMg_sequential_combined} in a similar format to Figure \ref{fig:Li_sequential_combined}.
However, each specific lithium-magnesium composition is sampled separately and in parallel for each active learning iteration.
Figure \ref{fig:LiMg_sequential_combined} summarizes each of them and SI Figure 7 shows the performance on the individual datasets.
We note that the random sampling was prematurely cutoff at 1600 samples due to compute constraints.

The trends of the errors, number of uncertain and high error samples in Figure \ref{fig:LiMg_sequential_combined}(a,b,d) are similar to the first test. 
However, with the high complexity of phase space, the precision, recall, and F1 calibration metrics remain much higher for a larger amount of the dataset size.
As compared to the first active learning test where the calibration metrics dropped to zero quickly, the calibration metrics for all the methods are nearly 0.5 throughout the active learning test.
This very clearly shows the generalizability and flexibility of the uncertainty cutoffs to maintain high precision and recall for diverse chemical space and data distributions.

A further difference between the two datasets is the presence of rare-events in the second active learning test. 
Since rare events are not well correlated to other samples, they are challenging for the machine learning model to learn and therefore have errors above the stability limit set in the benchmark.
The distribution of errors and forces for the five datasets is in SI Figure 6, showing a very different distribution for the three lithium-magnesium mixtures.
The effect of these rare events can be seen Figure \ref{fig:LiMg_sequential_combined}(f), where the two datasets without rare events are completely sampled in the first couple active learning iterations, but the rare events prevent sampling the entire datasets for many more iterations, even for the oracle model.

In the second test, we again see that the distance model has the closest data efficiency in Figure \ref{fig:LiMg_sequential_combined}(c) to the oracle followed by the Gaussian process and then the ensemble.
Coupling this with the lower number of uncertain and high error datapoints in Figure \ref{fig:LiMg_sequential_combined}(b,d) and the greater stability in Figure \ref{fig:LiMg_sequential_combined}(f), the distance model seems to perform the best on each of these tests.

\section{Discussion}

These benchmarks show the power of the statistical cutoff for use in complicated active learning settings.
The statistical definition of a cutoff for errors and uncertainties provides for the flexible use of different uncertainty methods reliably in active learning environments.
By understanding that MLIP errors like all deep learning model errors obey a statistical distribution, we can treat a given error based on it's likelihood to have been drawn from the theoretical distribution.
By applying this same understanding to the distribution of uncertainty predictions, we can find the likelihood that a given uncertainty prediction corresponds to the same distribution as the MLIP dataset.
In active learning, this allows us to correctly lable datapoints as uncertain even though the actual uncertainty predictor may not be well calibrated.
We show that by utilizing the statistical cutoffs, we can match the performance of the best uncertainty quantification method, which is utilizing the true error.
Furthermore, the statistical definition of both the error and uncertainty cutoff allows for a direct correspondence between the two allowing for an interpretable definition of the uncertainty cutoff.

The statistical cutoff improves over utilizing a fixed cutoff across all active learning methods as it provides a theoretical underpinning for when the sampling distribution is the same as the training or validation distributions.
Then, when the statistical cutoff identifies that the two distributions are aligned, we can make an informed choice about how to improve the true error since we have confidence that the validation error is similar to the sampling error.
For instance, the model error could be improved by either more data could be include into the training set by continually sampling or the model capacity could be increased to improve the representation power of the model.

One downside of the benchmarks is that by sampling the phase space at large intervals, there is the possibility for single datapoints to be poorly correlated to the rest of the dataset.
This becomes an issue since these highly uncorrelated samples will be labeled with high uncertainty and then added to the dataset.
The best case scenario is that this uncorrelated datapoint is added to the training set and the model trains to the rare events. Even in the best case scenario, the model may not train well due to poor representation in the dataset.
The worst case scenario is that if the uncorrelated point is split into the validation dataset and there are no other data points that are correlated with the structure resembling this high error structure, the value will remain a high error in the validation dataset and not be able to be improved upon with the other structures.
This points to improved sampling methods needing to be incorporated in production active learning environments, where the MLIP is utilized to directly sample the phase space.
However, we leave this for future work since relabeling sampled points with DFT is not suitable to the breadth of this work and will only be carried out with a single uncertainty quantifier. 

We also believe that there is a continued need to develop better benchmarking tasks for both uncertainty methods and active learning.
The benchmark needs to be chemically complex such that very data efficient equivariant models do not completely describe the dataset with few points.
However, the dataset also cannot contain rare events that are disjoint from the training distribution as seen in the $Li_3Mg$ and $LiMg_3$ force distributions in the SI.
Furthermore, the size of this dataset should be able to be reduced in size (N$<$100) for equivariant models to be quick to train on the dataset.
This will keep the cost of training MLIPs down such that multiple active learning runs can be run to average out the stochastic effect on each test.

\backmatter

\bmhead{Supplementary information}

Explanations of the benchmark datasets and descriptions of the uncertainty quantification methods with corresponding parity and calibration plots can be found in the supplementary information.

\bmhead{Acknowledgments}
This work was supported by the Defense Advanced Research Project Agency (DARPA) under contract no. HR00112220032. The results contained herein are those of the authors and should not be interpreted as necessarily representing the official policies or endorsements, either expressed or implied, of DARPA, or the U.S. Government.

\bibliography{references.bib}

\end{document}


\title[Article Title]{Supplementary Information: Statistical methods for resolving poor uncertainty quantification in machine learning interatomic potentials}


\author*[1]{\fnm{Emil} \sur{Annevelink}}\email{eannevel@andrew.cmu.edu}

\author*[1]{\fnm{Venkatasubramanian} \sur{Viswanathan}}\email{venkvis@cmu.edu}

\affil[1]{\orgdiv{Mechanical Engineering}, \orgname{Carnegie Mellon University}}

\maketitle

\section{Dataset Descriptions}

\begin{table*}[t]
\centering
\begin{tabular}{p{0.06\textwidth} | p{0.075\textwidth} | p{0.075\textwidth} | p{0.075\textwidth} | p{0.07\textwidth} | p{0.09\textwidth} | p{0.075\textwidth} | p{0.065\textwidth} | p{0.075\textwidth}}
   Index & Species & Lattice & N Atoms & Temp. (K) & Sampler & Length & Dump Freq. & Dataset Size  \\
    \hline
    \hline
    1 & Li & BCC & 250 & 600 & NVT & 10ns & 100ps & 101 \\
    \hline
    2 & Li & BCC & 250 & 600 & NVT & 10ns & 10ps & 1001 \\
    \hline
    3 & Li & BCC & 250 & 600 & NVT & 1ps & 10fs & 101 \\
    \hline
    \hline
    4 & Li & BCC & 125 & 600 & NPT & 1ps & 10fs & 101 \\
    \hline
    5 & Li & BCC & 125 & 600 & NPT & 10ns & 10ps & 1001 \\
    \hline
    6 & Li$_3$Mg & BCC & 125 & 600 & NPT & 1ps & 10fs & 101 \\
    \hline
    7 & Li$_3$Mg & BCC & 125 & 600 & NPT & 10ns & 10ps & 1001 \\
    \hline
    8 & LiMg & BCC & 125 & 600 & NPT & 1ps & 10fs & 101 \\
    \hline
    9 & LiMg & BCC & 125 & 600 & NPT & 10ns & 10ps & 1001 \\
    \hline
    10 & LiMg$_3$ & HCP & 61 & 600 & NPT & 1ps & 10fs & 101 \\
    \hline
    11 & LiMg$_3$ & HCP & 61 & 600 & NPT & 10ns & 10ps & 1001 \\
    \hline
    12 & Mg & HCP & 61 & 600 & NPT & 1ps & 10fs & 101 \\
    \hline
    13 & Mg & HCP & 61 & 600 & NPT & 10ns & 10ps & 1001 \\
\end{tabular}
\caption{Description of datasets used in the paper.}
\end{table*}

The thirteen datasets used in the main text can be separated into two main categories.
The first category is for a single component liquid, while the second is for a multi-component system spanning solid and liquid phases and BCC and HCP crystal lattices.
The datasets are constructured by sampling NVT or NPT ensembles using a classical potential for the given systems \cite{kimAtomisticModelingPure2012}.
After sampling, the configurations are recalculated with density functional theory using the GPAW package \cite{GPAW1,GPAW2}.
Convergence testing yielded the following GPAW parameters: 0.05 width methfessel-paxton occupation, h-spacing of 0.18, and k-point density of 30, which were used for all calculations.

The first type of dataset is meant to be complex enough to require active learning, but still relatively simple.
The lithium system was sampled at 600 Kelvin, where the lithium is above its melting temperature to give a diverse set of coordinates.
A cubic, 5x5x5 supercell was used for a total of 250 lithium atoms.
The second type of dataset aims to create a complex dataset to strain the uncertainty classification and the active learning.
The chemical complexity is increased from a single component lithium system to a dual lithium-magnesium system.
Furthermore, while 600K is still used, the system now includes both liquid and solid states as well as BCC and HCP crystal structures as the configurations are different for varying lithium-magnesium stochiometries.
Furthermore, three vacancies are introduced into each supercell.
The potential energy surface is sampled under an NPT ensemble using Nos\'e--Hoover thermo- and barostats to further diversify the atomic configurations.

\section{Description of Uncertainty Quantification Methods}

The most common uncertainty methods in the Machine Learning Interatomic Potential (MLIP) literature is using an ensemble of models \cite{zhangDPGENConcurrentLearning2020,wenUncertaintyQuantificationMolecular2020a,zhuFastUncertaintyEstimates2022}.
Recent effort has gone into using a single MLIP model with a separate uncertainty classifier such as a sparse Gaussian Process \cite{vandermauseOntheflyActiveLearning2020}.
Finally, distances in latent space have been shown to accurately predict the accuracy of chemical properties\cite{janetQuantitativeUncertaintyMetric2019}.
The following three methods describe the uncertainty methods utilized in the main text and implemented in the code base.

\subsection{Ensemble}

The distribution of predictions for an ensemble of models gives its prediction accuracy.
The invariant prediction from the models is given by

\begin{equation}
    \sigma_f = A_e \sqrt{\sum_{i \in xyz} \sum_{j \in n_e} \frac{f_{i,j}-\bar{f_i}}{n_e}}
\end{equation}
which gives the norm of the standard deviations of each force component for a given model ensemble size $n_e$.
A prefactor is used to calibrate the ensemble prediction to the errors in the validation dataset using the negative log-likelihood, where $\sigma_f$ corresponds to the standard deviation of a Maxwell distribution.

\subsection{Gaussian Process}

The Gaussian Process uncertainty uses the `gpytorch'\cite{gardner2018gpytorch} package to create a sparse Gaussian Process model to reproduce the datapoints.
The sparse Gaussian process model is utilized according to prior work done in Flare \cite{vandermauseOntheflyActiveLearning2020}, which shows the poor scaling of an exact GP.
The inducing points of the sparse GP model are initialized through clustering the MLIP latent features using k-means and then optimized for 1000 epochs.

\subsection{Distance}

The latent distance uncertainty metric is inspired by Janet et al. \cite{janetQuantitativeUncertaintyMetric2019} and adjusted for quantifying the uncertainty from MLIPs.
The uncertainty is given by 

\begin{equation}
    \sigma_f = \sigma_0 + \sum_{i \in n_l} | \sigma_i * d_i | \hspace{0.01in},
\end{equation}
where the invariant uncertainty prediction $\sigma_f$ is the sum of a baseline uncertainty $\sigma_0$ and a linear term for each component of the $n_l$ dimensional latent space.
The distance $d_i$ is the minimum distance between a sample point and the latent features of the training set.

\section{Parity Plots}

\begin{figure*}[h]
    \centering
    \includegraphics[width=0.9\textwidth]{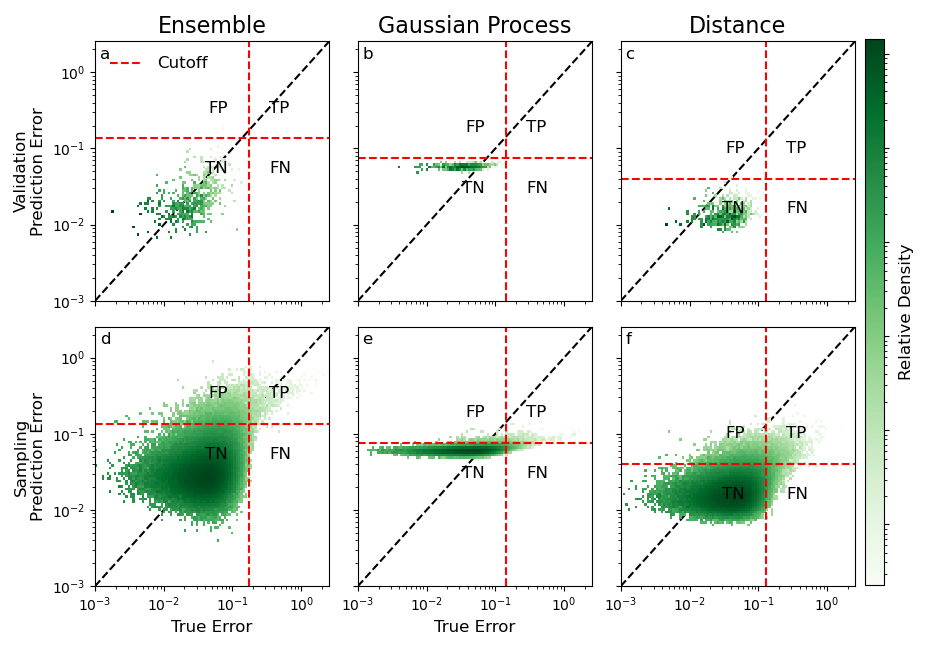}
    \caption{Parity plots of each uncertainty models on the validation and sampling datasets.}
    \label{fig:parity}
\end{figure*}

The statistical cutoff is a method of understanding the accuracy of the uncertainty quantification tools.
SI Figure \ref{fig:parity} shows the parity plots of the uncertainty quantification methods for the models trained during the first active learning iteration for the combined lithium dataset for both the validation and sampling datasets.
The parity plots identify the challenge that uncertainty quantification techniques need to address, which is the ability to accurately predict the uncertainty that is trained on only a single class.
For example, all the validation datapoints are correctly labelled as certain meaning that all the validation labels are true negatives.
During production, the uncertainty quantifier therefore needs to identify true positives based solely on the true negatives in the validation dataset.
This is the core issue for why developing uncertainty models for MLIPs has been so challenging.
However, the uncertainty cutoff can be used to address the difficulty in training accurate uncertainty models.

Like we see with the histograms in Main Text Figure 1 and 2, the statistical thresholds separate the high density regions from the low density regions.
However, due to the imperfect uncertainty prediction and the different methods of calibrating each uncertainty technique, the uncertainty cutoffs are quite different even though the error cutoffs are nearly identical.
The flexible nature of the uncertainty cutoff enables the use of different uncertainty methods and provides a direct connection to the error distribution to create the decision boundary.

The performance of the different methods can be visually inspected based on the shapes of the 2d histograms.
The range of the ensemble sampling uncertainty predictions is most similar to the error range followed by the distance and then finally by the Gaussian Process model.
Most of this has to do with the uncertainty prediction definition for each model and the training paradigm utilized.
However, part of it also has to do with the actual models themselves and their limitations.
For example, for a point far away from the training distribution, the Gaussian Process model will plateau further than the largest lengthscale for the best-fit radial distribution function.
This is contrary to the distance based metric that will monotonically increase for latent features far from the training distribution meaning that the Gaussian Process predictions will always have a narrower range than the distance metric.

\newpage
\section{Calibration Plots}

Calibration curves are an alternative method of checking the accuracy of the uncertainty predictions.
Since the negative log-likelihood is utilized to train each of the uncertainty methods, the uncertainty prediction is a standard deviation.
The calibration curve are plot according to Tran et al. \cite{tranMethodsComparingUncertainty2020}, where the predicted cumulative density function is found by normalizing the error by the predicted standard deviation.
The resulting calibration curves and the area between the calibration curve and the x=y ideal line implies the ability of the uncertainty models to best predict the data.

The ensemble model has the lowest miscalibration area for both the validation and sampling datasets.
Interestingly, the ensemble miscalibration area is lower for the sampling dataset compared to the validation dataset.
As can be seen in the parity plots in Figure \ref{fig:parity}(d), the ensemble prediction bends over slightly below the x=y line meaning the large uncertainty predictions are overconfident.
The Gaussian Process and Distance Models, although they look well calibrated on the parity plot are systematically under- and over-confident respectively.

\begin{figure*}[h]
    \centering
    \includegraphics[width=0.9\textwidth]{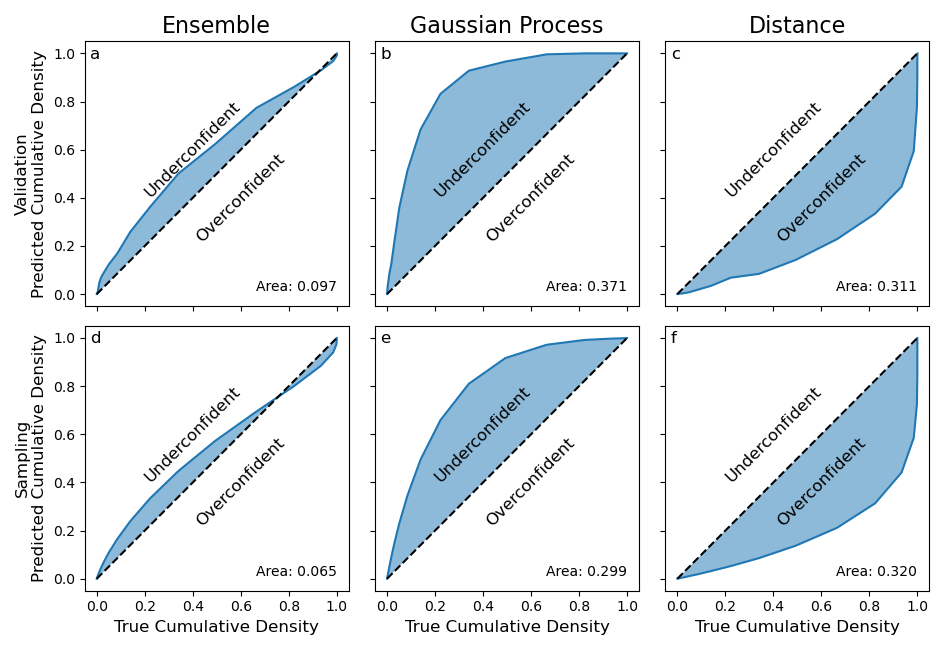}
    \caption{Calibration curves of each uncertainty models on the validation and sampling datasets.}
    \label{fig:calibration}
\end{figure*}

\clearpage
\section{Precision, Recall, F1}

The precision and recall are presented along with the F1 scores that are in the main text.
The precision and recall plots show the contributions to the F1 score.
In each of the below figures, the error cutoff matches where the precision goes to zero for low uncertainties, while the uncertainty cutoff matches the threshold where the recall goes to zero for low errors.
The separated precision and recall curves further support how the statistical cutoffs separate the MLIP errors and uncertainty predictions into different regions of phase space and how when used together separate high and low F1 score regions

\begin{figure*}[h]
    \centering
    \includegraphics[width=0.9\textwidth]{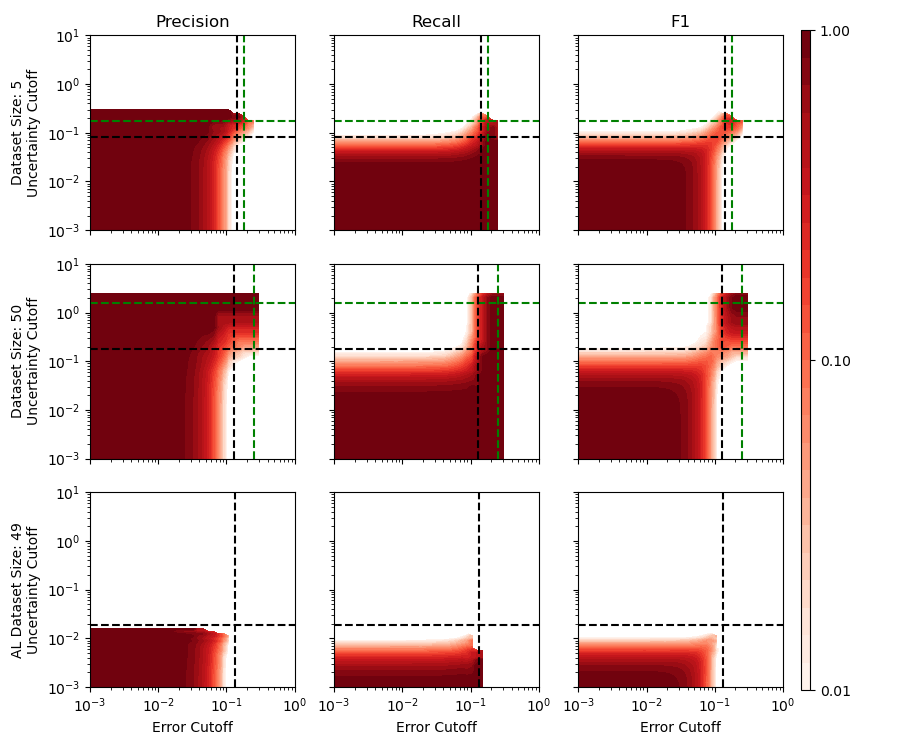}
    \caption{The precision, recall, and F1 scores for a sweep of error and uncertainty cutoffs using the uncertainty predicted from the standard deviation of an ensemble of models. The first row corresponds to 5 randomly sampled datapoints, the second to 50, and the third row to an active learning dataset with a final dataset size of 49.}
    \label{fig:prec_rec_f1_ens}
\end{figure*}

\begin{figure*}[h]
    \centering
    \includegraphics[width=0.9\textwidth]{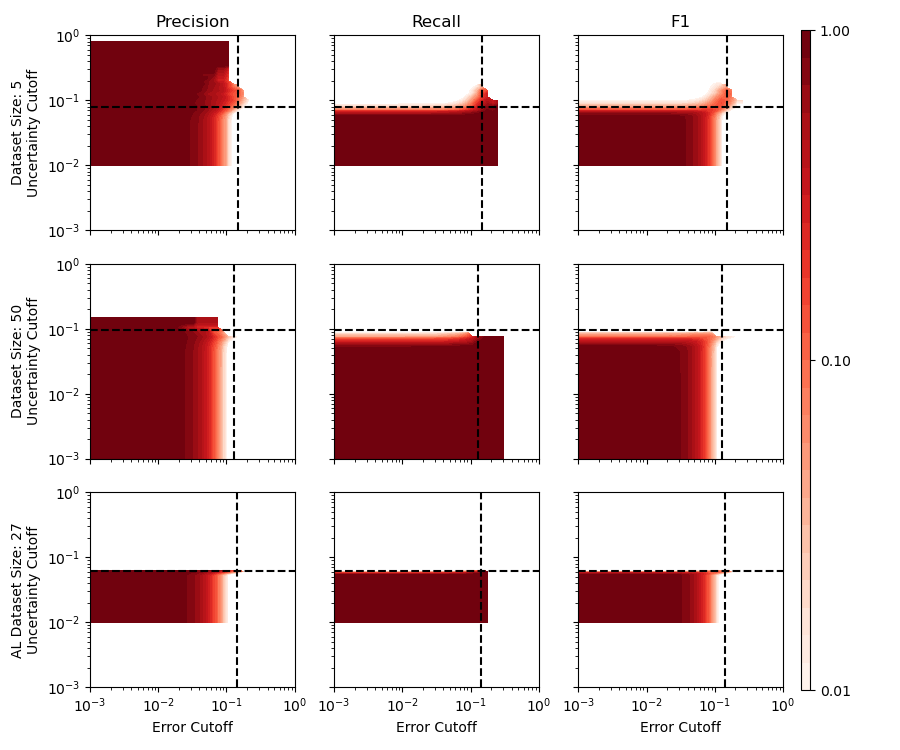}
    \caption{The precision, recall, and F1 scores for a sweep of error and uncertainty cutoffs using the uncertainty predicted from a Gaussian Process model. The first row corresponds to 5 randomly sampled datapoints, the second to 50, and the third row to an active learning dataset with a final dataset size of 27.}
    \label{fig:prec_rec_f1_GPR}
\end{figure*}

\begin{figure*}[h]
    \centering
    \includegraphics[width=0.9\textwidth]{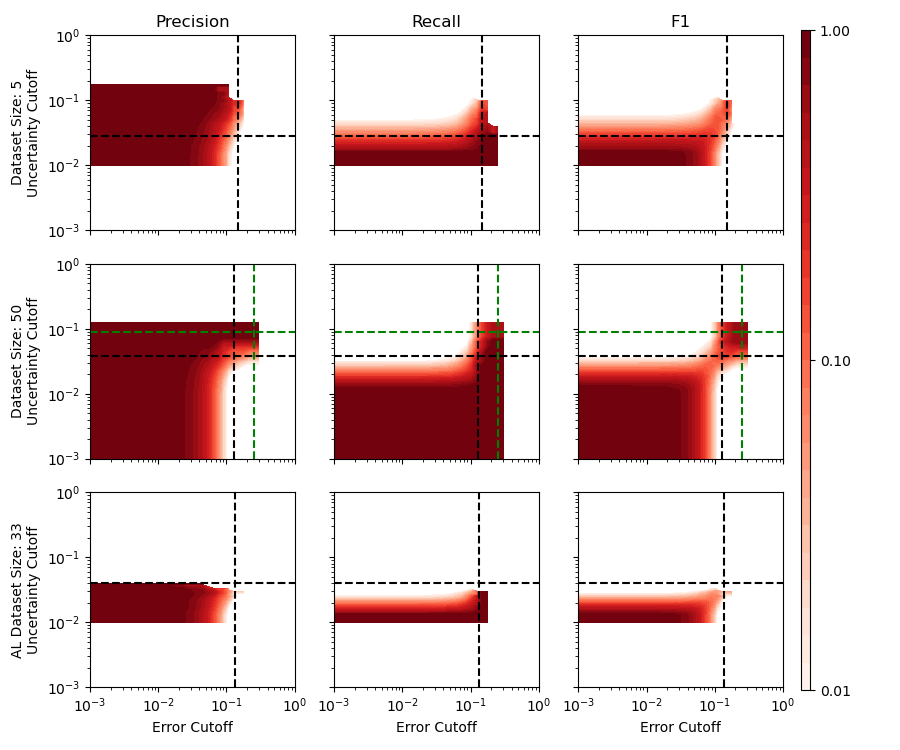}
    \caption{The precision, recall, and F1 scores for a sweep of error and uncertainty cutoffs using the uncertainty predicted from a latent distance model. The first row corresponds to 5 randomly sampled datapoints, the second to 50, and the third row to an active learning dataset with a final dataset size of 33.}
    \label{fig:prec_rec_f1_dis}
\end{figure*}

\clearpage
\newpage
\section{Lithium - Magnesium Dataset Energy and Force Distributions }

\begin{figure}[h]
    \centering
    \includegraphics[width=\textwidth]{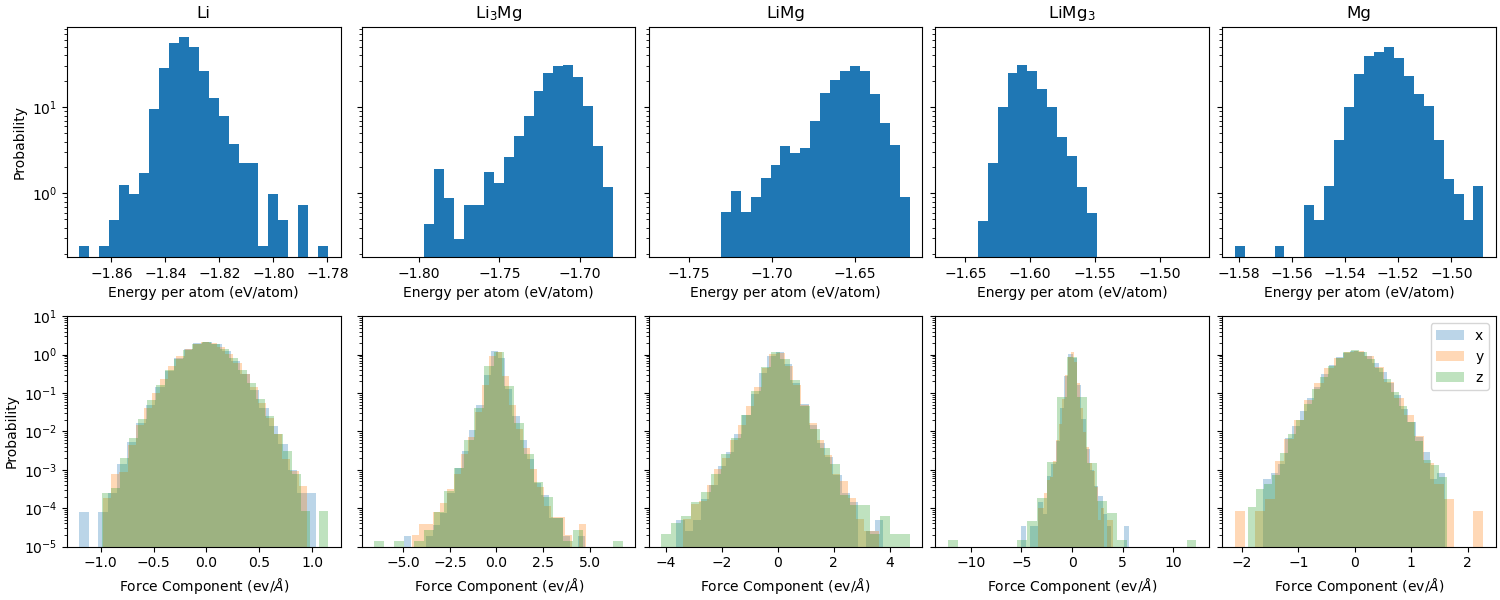}
    \caption{Energy and force distributions for the datasets used in the second active learning benchmark.}
    \label{fig:LiMg-energies-forces}
\end{figure}

The complexity of the Li-Mg datasets is seen by the different energy and force distributions in Figure \ref{fig:LiMg-energies-forces}.
The energy distributions are in the first row, while the force component distributions are in the second row. 
The energy distributions for the pure component Li and Mg systems are relatively symmetric, but the alloy datasets have a noticeable skew to the datasets.
Further, while the force component datasets are all symmetric, the expected Gaussian distribution is only present for the pure component systems.
For the alloy systems, the tails are much longer due to the presence of rare events from the self-diffusion of atoms allowed due to the presence of vacancies. 

\newpage
\section{Benchmark 2 active learning performance separated by chemical space}

\begin{figure}[h]
    \centering
    \includegraphics[width=\textwidth]{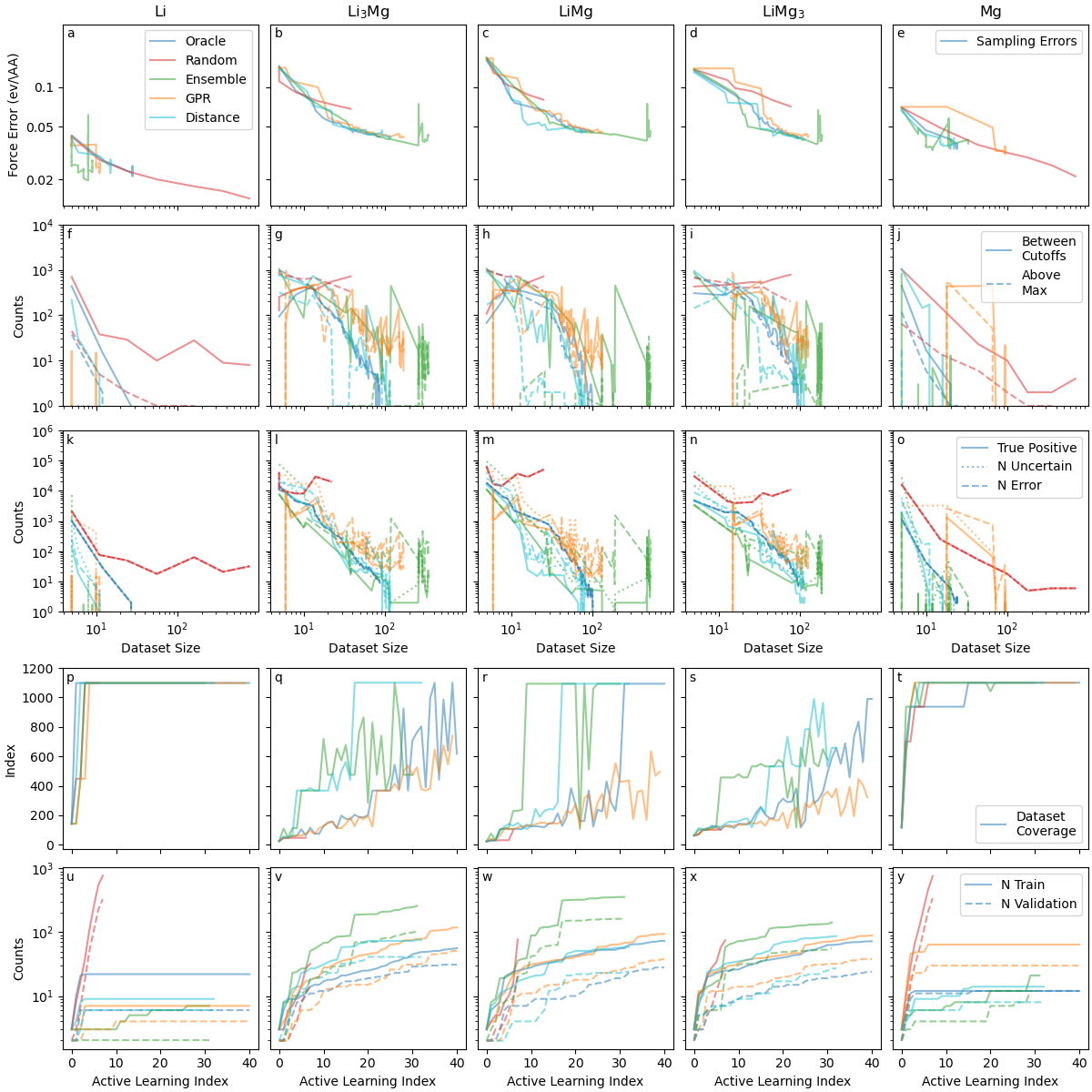}
    \caption{Performance of active learning benchmark 2 separated by dataset.}
    \label{fig:LiMg-performance-separate}
\end{figure}

The active learning performance for the second benchmark is reported in Figure \ref{fig:LiMg-performance-separate}.
The individual performance metrics show that the active learning has a much harder time with the three alloy datasets in the center three columns.
In particular, while the entire datasets are spanned quickly for the two pure component datasets in Figure \ref{fig:LiMg-performance-separate}(p,r), the dataset index increases slowly due to the presence of rare events as shown in Figure \ref{fig:LiMg-energies-forces}.

\bibliography{references.bib}